\documentclass[twocolumn,superscriptaddress,epsf,psfig]{revtex4-1}
\usepackage{amssymb}
\usepackage{epsfig}
\DeclareMathAlphabet{\pazocal}{OMS}{zplm}{m}{n}            
\usepackage{graphicx}
\usepackage{color}
\usepackage{bm}
\usepackage[normalem]{ulem}     
\usepackage{soul}
\usepackage{hyperref}
\usepackage{amsmath}
\usepackage{braket} 
\usepackage{rotating}
\usepackage{multirow}

\begin{document}
\title{Anti-symmetric Compton scattering in LiNiPO$_4$: Towards a direct probe of the magneto-electric multipole moment}    
\author{Sayantika Bhowal$^*$}
\affiliation{Materials Theory, ETH Zurich, Wolfgang-Pauli-Strasse 27, 8093 Zurich, Switzerland}  
\author{Daniel O'Neill$^*$}
\affiliation{Department of Physics, University of Warwick, Coventry, CV4 7AL, UK}
\author{Michael Fechner$^*$} 
\affiliation{Max Planck Institute for the Structure and Dynamics of Matter, 22761 Hamburg, Germany}  
\author{Nicola A. Spaldin}
\affiliation{Materials Theory, ETH Zurich, Wolfgang-Pauli-Strasse 27, 8093 Zurich, Switzerland}
\author{Urs Staub}
\affiliation{Swiss Light Source, Paul Scherrer Institute, 5232 Villigen PSI, Switzerland}
\author{Jon Duffy}
\affiliation{Department of Physics, University of Warwick, Coventry, CV4 7AL, UK}
\author{Stephen P. Collins}
\affiliation{Diamond Light Source Ltd, Diamond House, Harwell Science \& Innovation Campus, Didcot, Oxfordshire, OX11 0DE}

\date{\today}

\begin{abstract}
 We present a combined theoretical and experimental investigation of the anti-symmetric Compton profile in LiNiPO$_4$ as a possible probe for magneto-electric toroidal moments. Understanding as well as detecting such magneto-electric multipoles is an active area of research in condensed matter physics. Our calculations, based on density functional theory, indicate an anti-symmetric Compton profile in the direction of the $t_y$ toroidal moment in momentum space, with the computed anti-symmetric profile around four orders of magnitude smaller than the total profile. The difference signal that we measure is consistent with the computed profile, but of the same order of magnitude as the statistical errors and systematic uncertainties of the experiment. Our results motivate  further  theoretical work to understand the factors that influence the size of the anti-symmetric Compton profile, and to identify materials exhibiting larger effects. \\
\vspace*{0.2cm} 

\noindent 
 $^*$These authors contributed equally to this work.

\end{abstract}

\maketitle

\section{Introduction}  \label{sec0}

Magneto-electric (ME) multipoles are key to understanding the linear ME response in solids, in which an applied electric field induces a linear order magnetization, and vice versa.
In particular, the second-rank ME multipole tensor, defined as ${\cal M}_{ij} =  \int  r_i \mu_j (\vec r) d^3r$ \cite{Spaldin2008,Spaldin2013}, where $\vec \mu(\vec r)$ is the magnetization density, has the same symmetry as the linear ME response tensor. Both are only non-zero when space-inversion (${\cal I}$) and time-reversal  (${\cal T}$) symmetries are broken simultaneously, and there is a one-to-one correlation between their components.  For example, materials with an antisymmetric off-diagonal linear ME response also have non-zero antisymmetric off-diagonal elements in their ${\cal M}_{ij}$ tensor \cite{Spaldin_2021}. 

The ME multipole ${\cal M}_{ij}$ tensor  can be decomposed into three irreducible (IR) components,  the ME monopole, ME dipole (toroidal) moment, and ME quadrupole moment, as summarized in Table \ref{tab1} \cite{Spaldin2008,Spaldin2013}. In the present work, we are  particularly interested in the ME toroidal moment $\vec t =  \frac{1}{2} \int  \vec r \times \vec \mu (\vec r) d^3r$, the components of which form the  
the antisymmetric off-diagonal elements of the multipole tensor ${\cal M}_{ij}$, 
\begin{eqnarray}  \label{MEt}
t_i= \frac{1}{2} \varepsilon_{ijk} {\cal M}_{jk}  \quad.
\end{eqnarray}
The ME toroidal moment has been proposed as the order parameter for a form of hidden ferroic order, known as \emph{ ferrotoroidic}, to complete the set of primary ferroics with the 
 existing established ferromagnetism, ferroelectricity, and ferroelasticity \cite{Spaldin2008,Schmid,Aken}. This proposal motivated considerable interest in ME toroidal moments in solids, leading to experimental efforts to detect them using resonant x-ray diffraction \cite{Arima2005,Staub2009,Lovesey,Staub2010,Staub}, magneto chiral dichroism \cite{Kubota,Sessoli}, and optical measurements \cite{Jung2004}, as well as to image ferrotoroidic domains \cite{Aken}. While a theory of  \emph{toroidization} (toroidal moment per unit volume) in periodic solids has been developed \cite{Claude2007}, a direct and quantifiable link of detected signals from toroidal moments to the underlying electronic structure is still lacking.

 Recently, the occurrence of an antisymmetric component in the Compton scattering profile, which measures the electron momentum density $\rho (\vec p)$, was proposed as a possible direct probe of the ME toroidal moment \cite{Collins2016}. For materials that are symmetric in  either or both  of ${\cal I}$ and ${\cal T}$, the Compton profile, $J(p)$, which is a projection of electron momentum density of the form $J (p_z) = \int \rho(\vec p) dp_xdp_y$, is symmetric in momentum space, because under both $\cal T$ and $\cal I$, $\vec p \rightarrow -\vec p$. For materials that lack {\it both} ${\cal I}$ and ${\cal T}$  symmetries, however, an antisymmetric contribution to the Compton profile is allowed, suggesting Compton scattering as a sensitive probe of atomic-scale magnetoelectric properties.
A first investigation was made using magnetoelectric GaFeO$_3$ \cite{Collins2016}, for which a density functional study of the ideal material predicted a measurable asymmetry. The measured Compton asymmetry, however, was within the experimental uncertainty, possibly due to the well-known inter-site mixing of Ga and Fe in GaFeO$_3$, and so no clear assignment of a toroidal moment could be made. 

\begin{table*} [t]
\caption{ The three irreducible components of the ME multipole tensor ${\cal M}_{ij}$, that contribute to the second order term ${\cal E}^{(2)}_{int}$ in the multipole expansion of the interaction energy in presence of an external magnetic field $\vec H$.
}
\setlength{\tabcolsep}{2pt}
\centering
\begin{tabular}{ c |c |c| c }
\hline
ME multipole & ME monopole ($a$) & ME toroidal moment ($\vec t$) & ME quadrupole moment ($q_{ij}$) \\
\hline\hline
Definition &     $a= \frac{1}{3} {\cal M}_{ii} =  \int  \vec r \cdot \vec \mu (\vec r) d^3r$                                     & $t_i= \frac{1}{2} \varepsilon_{ijk} {\cal M}_{jk} =  \frac{1}{2} \int  \vec r \times \vec \mu (\vec r) d^3r$  & $q_{ij} =  \frac{1}{2}   \int  \big(r_i \mu_j + r_j \mu_i - \frac{2}{3} \delta_{ij} \vec r \cdot \vec \mu \big)  d^3r $ \\ [1ex]
Rank        & 0 (scalar)		                                  & 1 (vector)	 & 2 (symmetric traceless tensor)\\ [1ex]
 ${\cal E}^{(2)}_{int} =-  \int  r_i \mu_j (\vec r) \partial_i H_j (0) d^3r$ & $-a (\vec \nabla \cdot \vec H)_{\vec r =0}$ & $-\vec t \cdot (\vec \nabla \times \vec H)_{\vec r =0}$ & $q_{ij}(\partial_iH_j+\partial_jH_i)$\\ [1ex]
  \hline
  \end{tabular}
\label{tab1}
\end{table*}

Here, we present a combined theoretical and experimental Compton scattering study of a magnetoelectric  material, lithium nickel phosphate, that has no reported tendency to site disorder. LiNiPO$_4$ is an anti-ferromagnetic insulator, which shows an off-diagonal linear ME response  \cite{Mercier,Jensen,Exp2020} and hosts a ME toroidal moment \cite{Spaldin2013} in its magnetic ground state. (Note that many magnetic-field induced magnetic transitions have recently been identified, resulting in other phases with linear and quadratic ME effects which we do not treat here \cite{Exp2020}.)
Our calculations using density functional theory (DFT) indeed show the presence of an anti-symmetric Compton profile along the $y$ direction in momentum space, consistent with the presence of the ME toroidal moment component $t_y$ in LiNiPO$_4$. The calculated intensity of the anti-symmetric part of the profile is about four orders of magnitude smaller than that of the corresponding total profile.  Our measured anti-symmetric signal  is  of the same order of magnitude as our computed value, but is also of the same order of magnitude as the statistical errors and systematic uncertainties of the experiments. Our main finding, therefore, is that, while an antisymmetric Compton scattering would indeed indicate the existence of a magnetoelectric multipole in a material, improvements in the experimental sensitivity and/or identification of materials with a larger response will be necessary for an unambiguous determination.

We organize the paper as follows. In section \ref{sec1}, we provide a detailed description of the crystal structure of LiNiPO$_4$ and discuss the theoretical and experimental methods used in the present work to analyze its Compton profile. This is followed by the discussion of our results in section \ref{sec2}. Here, we first discuss the density functional results and the experimental results individually and, then, compare our theory and experiments in order to gain insight into the measured Compton profile in LiNiPO$_4$. We also discuss the possible roles of domain averaging and convolution of the profile in reducing the anti-symmetric signal in the measurements. Finally, we summarize our findings in section \ref{sec3}.

\section{Crystal structure and Methods} \label{sec1}

In this section, we discuss the crystal structure of LiNiPO$_4$, followed by the theoretical and experimental methods that were employed in the present work to obtain and study the Compton profile of the material.

\subsection{Crystal Structure}

\begin{figure}[t]
\centering
\includegraphics[scale=.3]{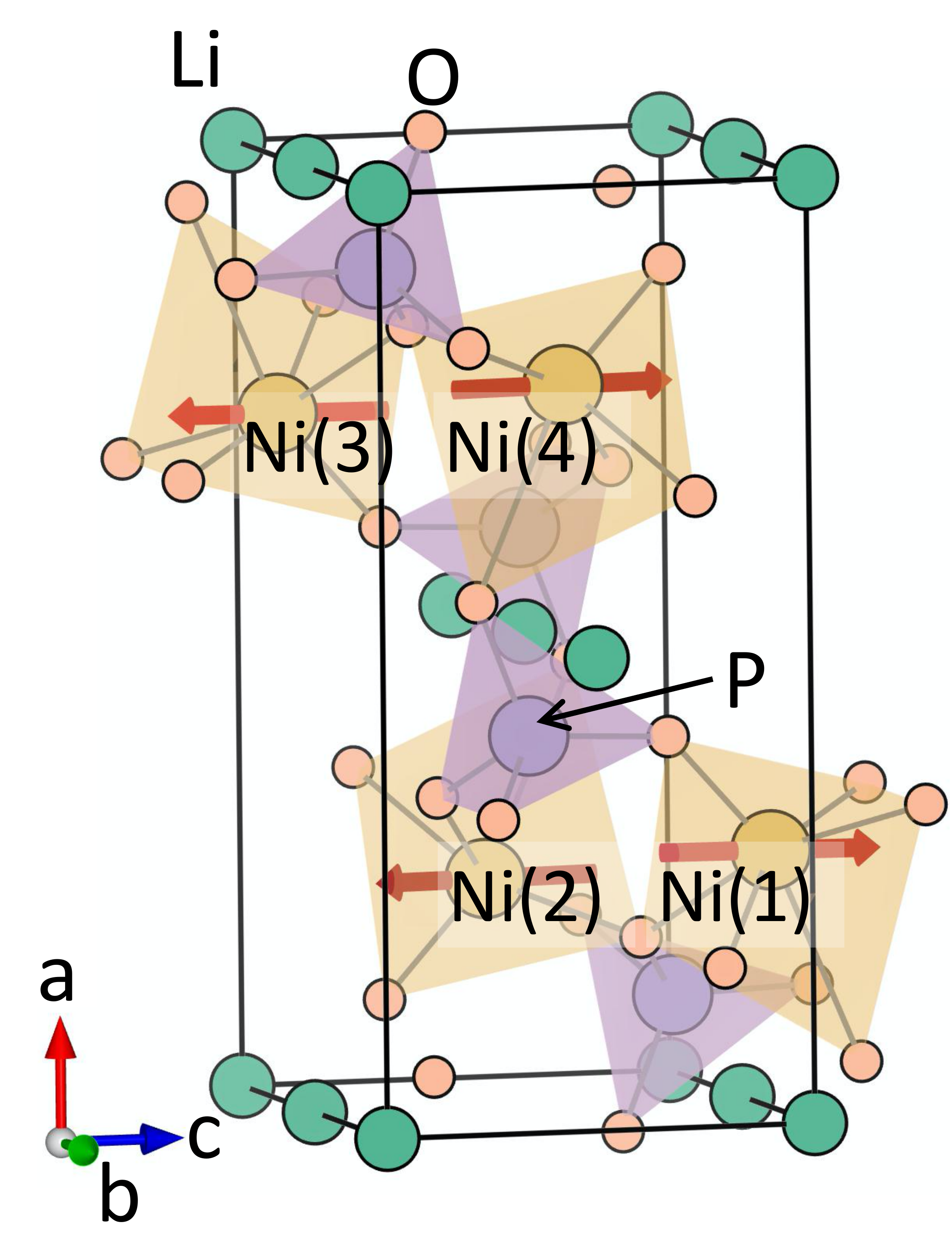}
 \caption{The crystal and magnetic structure of LiNiPO$_4$. The red arrows denote the spin moment at the Ni sites corresponding to the magnetic ground state with $Pnm'a$ symmetry.
 }
 \label{fig1}
 \end{figure}

LiNiPO$_4$ crystallizes in the orthorhombic $Pnma$ structure, with the point group symmetry $D_{2h}$ \cite{Abrahams}. The unit-cell structure is shown in Fig. \ref{fig1}. As seen from this figure, the crystal structure consists of distorted NiO$_6$ octahedra, which are connected to each other by PO$_4$ tetrahedral units. The unit cell consists of four formula units, with the four Ni atoms occupying the Wyckoff positions $4c$. In contrast to the previously studied GaFeO$_3$, the crystal structure of LiNiPO$_4$ has inversion $\cal I$ symmetry in the absence of magnetic ordering. The magnetic arrangement at the Ni sites in the ground state breaks both $\cal I$ and time-reversal $\cal T$ symmetries, leading to a linear ME effect and a non-zero toroidal moment. 
The combined $\cal I \cal T$ symmetry remains preserved in this magnetic ground state.

\subsection{Computational methods}

The electronic structure, the anti-symmetric Compton profile, and the ME multipoles for LiNiPO$_4$ are obtained using the linearized augmented plane wave (LAPW) method of density functional theory
as implemented in the ELK  code \cite{code,elk}. Spin-orbit coupling (SOC) is included explicitly in the calculations. A local Hubbard $U$ correction of $U_{\rm eff} =U -J = 4.25$ eV 
is applied to the Ni $3d$ electrons, within the LDA+SOC+$U$ formalism. 
In our calculation, we consider Li$:1s^22s^1$, Ni$:3p^63d^84s^2$, P$:3s^23p^3$, and O $:2s^22p^4$ electrons as valence electrons. 
The muffin-tin radii for Li, Ni, P, and O are taken to be 2.0, 2.4, 2.2, and 1.8 a.u. respectively.
In order to achieve self consistency, 
we use a basis set of $l_{max(apw)} = 8$, we sample the Brillouin zone with a $3\times6\times6$ k-point mesh, and take the product of the muffin-tin radius and the maximum reciprocal lattice vector to be 7. The calculations are carried out using the reported relaxed atomic positions calculated in LiNiPO$_4$ \cite{Spaldin2013}.

Once self-consistency is achieved, we compute the electron momentum density, which is further  
projected onto the selected momentum directions ($\vec p$) in order to obtain the desired Compton profile $J(\vec p)$ \cite{elk}. Next, we separate out the computed profile into symmetric $J^s (\vec p)$ and anti-symmetric $J^a (\vec p)$ parts using the following relation,
\begin{eqnarray} \nonumber \label{Jas}
J (\vec p) &=& 2^{-1} [J (\vec p)+J (-\vec p)] + 2^{-1} [J (\vec p) - J (-\vec p)]  \\
&=& J^s (\vec p) + J^a (\vec p)
\end{eqnarray}
Additional calculations with a denser $5\times10\times10$ k-point mesh confirm the convergence of each of the parts, $J^s (\vec p)$ and $J^a (\vec p)$. We normalize both profiles to the total number of valence electrons per formula unit of LiNiPO$_4$, 48, and add the isotropic core contribution to the symmetric part to obtain the total calculated Compton profile, which should be comparable with the measured profile. The core contribution is obtained from the 
Hartree-Fock calculations
of Biggs {\it et. al.} \cite{Biggs}.
To compare the computed anti-symmetric Compton profile with the measured value, we further convolute the computed profile with a Gaussian of full-width at half maximum (FWHM) = 0.44 a.u. to  mimic the experimental momentum resolution, which is dominated by the
detector energy resolution of 
$\Delta E/E \sim 7\times10^{-3}$.

We calculate the ME multipoles in the sphere around the Ni sites (referred to as the atomic-site contributions in Refs.~\onlinecite{Spaldin2013,Claude2007}) by decomposing the density matrix $\rho_{lm,l'm'}$ as described in Ref.~\cite{Spaldin2013}. The parity-odd tensor moments have contributions only from the odd $l-l'$ terms, i.e., $l$ must be different from $l'$; for the case of Ni with its valence $s$, $p$ and $d$ electrons, this means that we include the $s-p$ and $p-d$ matrix element contributions.

\begin{figure}[t]
\centering
\includegraphics[width=\columnwidth]{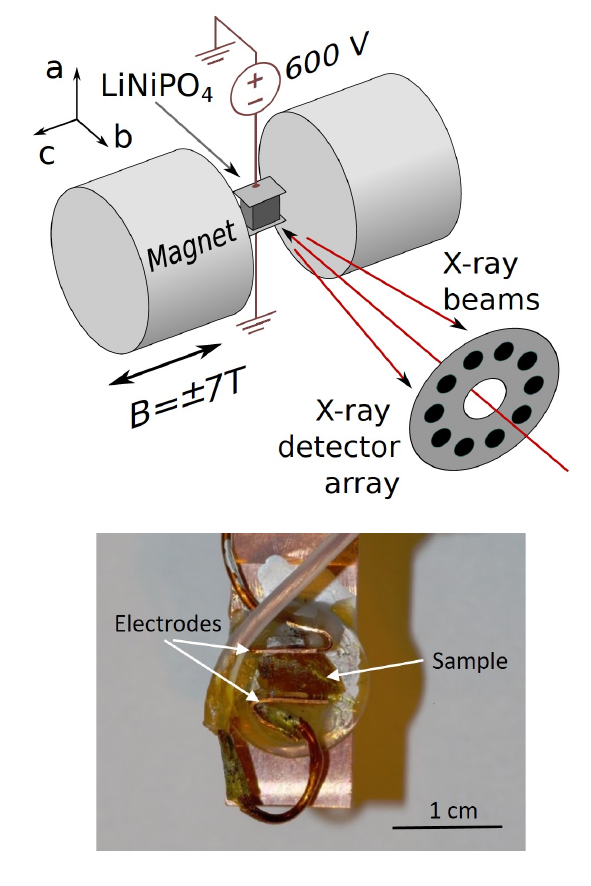}
 \caption{Experimental set up used to measure the anti-symmetric Compton profile in LiNiPO$_4$. The upper panel shows a schematic of the experimental set up, with $a$, $b$ and $c$ indicating the orientations of the crystal axes. The polar $a$ axis was oriented vertically in the setup. A reversible magnetic field was applied along the crystal $c$ axis. The $b$ axis, which is the orientation of the $t_y$ toroidal moment, was aligned close to the
direction of photon momentum transfer, which defines the projection
direction for the electron momentum density. The sample temperature was maintained at T$\sim$10K by the variable temperature stage of a helium-cooled superconducting magnet. The incident, linearly polarized x-ray beam of energy 184.3 keV passed through the aperture of a 10-element Ge detector,
which subsequently measured the energy spectrum of the Compton scattered x-rays close to back-scattering. The lower panel is a photograph of the crystal on the sample holder with electrodes attached perpendicular to the $a$-axis (vertical in the picture). 
 }
 \label{fig5}
 \end{figure}
\subsection{Experimental techniques}
Our experiment, carried out on BL08W(A) at SPring-8, utilized the Warwick University superconducting magnet to provide a 7T field perpendicular to the beam, coupled with a high voltage (600V), applied perpendicular to the beam and magnetic field. The experimental set up is shown schematically in Fig. \ref{fig5}. The sample was cooled through the ordering phase transition to around T=10K, while applying the magnetic and electric fields simultaneously. Then the cycle was repeated in an + - - + magnetic field sequence, with the same electric field but opposite magnetic field. Field cooling with crossed electric and magnetic fields in this geometry should produce a single magnetoelectric domain state with its net magnetoelectric toroidal moment along the beam direction, as required for our experiment. Any incomplete domain alignment would result in a reduction in the signal.

During the experiment, the greatest practical challenge was arcing due to the high voltage and low pressure helium exchange gas environment. We were therefore only able to provide a high enough electric field for one polarity and with the magnetic field applied. We were then able to reverse the magnetoelectric domains by ramping the magnetic field, although this required long acquisition times of one hour per cycle period. During this time even small drifts in detector gain can lead to parasitic signals that had to be treated very carefully. 

The energy spectrum of the measured inelastic scattering signal is in general easily mapped onto momentum space since the double differential cross-section is almost directly proportional to the Compton profile $J(p_z)$ (see, for example, Ref.~\cite{Cooper1985}).
Here, we were interested primarily in the antisymmetric difference profile, obtained by subtracting data sets measured with reversed magnetoelectric domains. Since there is a symmetry requirement for the two profiles to have the same area, the datasets were normalized prior to subtraction to ensure that this requirement was satisfied.

\section{Results and Discussion} \label{sec2}

 \begin{table*} [t]
\caption{The computed magnitudes and relative signs of ME multipoles and magnetic moments at the Ni site (Wyckoff position $4c$) corresponding to the $Pnm'a$ magnetic structure. The Ni(1), Ni(2), Ni(3), and Ni(4) atoms are indicated in Fig. \ref{fig1}.  
}
\setlength{\tabcolsep}{3pt}
\centering
\begin{tabular}{ c c c |c c c }
\hline
ME multipoles & Magnitude  & Relative orientation at  & Magnetic & Magnitude  & Relative orientation at  \\
                      & ($10^{-3} \mu_B$ a.u.) & Ni(1)-Ni(2)-Ni(3)-Ni(4)       & moment & $\mu_B$ & Ni(1)-Ni(2)-Ni(3)-Ni(4) \\  
\hline\hline
$t_y$ &  5.1  &  -    -   -   -  &  $m_x$& 0.01 & + + - -\\[1ex]     
$q_{xz}$ & 8.2 & -    -   -   - & $m_y$ & 0 & \\[1ex]    
$a$ & 5.7 & +  -  +   -  & $m_z$ & 1.69  & + - - +\\[1ex]    
$q_{x^2-y^2}$ & 0.3 & +  -  +  - &  &  &\\[1ex]  
$q_{z^2}$ & 5.9 & +   -  +  - &  &  &\\[1ex]  
  \hline
  \end{tabular}
\label{tab2}
\end{table*}
\subsection{Density-functional theory results} 

Our calculated lowest-energy  magnetic structure of LiNiPO$_4$ has magnetic space group $Pnm'a$, consistent with experimental measurements. In this magnetic configuration, the magnetic moments at the Ni sites are antiferromagnetically arranged, with computed spin moment of 1.69 $\mu_B$ at each Ni site and zero net magnetization. The Ni spins are primarily oriented along the $z$ direction with a small component along the $x$ direction at each site, as a result of a small antiferromagnetic canting, as shown in Fig. \ref{fig1}.

The magnetic  $Pnm'a$ symmetry of the ground state allows for various ME multipoles at the Ni sites, which have magnetic point group symmetry $mm'm$. These ME multipoles in turn have either ferro or anti-ferro type arrangements, as was pointed out previously in Ref.~\cite{Spaldin2013}. In particular, the symmetry allows for toroidal moments along the $y$ direction ($t_y$) and $q_{xz}$ quadrupole moment components, both with ferro-type arrangements, leading to a net $t_y$ and $q_{xz}$ in the system. In addition, the ME monopole $a$ and the $q_{x^2-y^2}$, $q_{z^2}$ quadrupole moments are also allowed at the Ni sites. These multipoles have opposite signs at the neighboring Ni sites, however, leading to net zero contributions to $a$, $q_{x^2-y^2}$, and $q_{z^2}$. Our density functional results for the ME multipoles are consistent with this symmetry analysis.  
The computed magnitudes of atomic-site contributions to the various allowed ME multipoles, and their relative orientations at different Ni-sites are listed in Table \ref{tab2}. As our calculation shows the presence of a net toroidal moment along the $y$ direction, we expect an anti-symmetric Compton profile along the same direction in momentum space \cite{Collins2016}, which we now proceed to discuss.

\begin{figure}[t]
\centering
\includegraphics[width=\columnwidth]{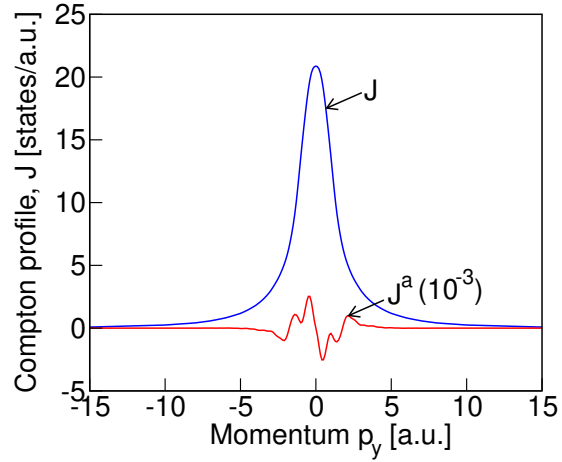}
 \caption{The computed (without convolution) total, $J$, and antisymmetric, $J^a$, Compton profile for the $Pnm'a$  magnetic structure (see Fig. \ref{fig1} for the spin arrangement) of LiNiPO$_4$.}
 \label{fig2}
 \end{figure}
\begin{figure}[t]
\centering
\includegraphics[width=\columnwidth]{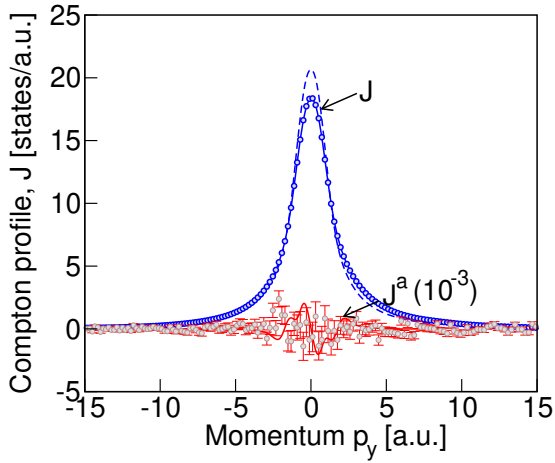}
 \caption{The measured Compton profile for LiNiPO$_4$ compared with the calculated profile of Fig.~\ref{fig2}. The measured total Compton profile is shown by the blue circle data points and the blue solid line, while the measured anti-symmetric profile, magnified by a factor of $10^3$, is indicated by the red circle data points and red vertical line error bars.  
 The convoluted computed profiles with a Gaussian of FWHM = 0.44 a.u. from Fig.~\ref{fig2} are shown as the blue dashed line (total profile) and red solid line (anti-symmetric part of the calculated profile, magnified by a factor of $10^3$). While the peak in computed total profile differs from the measured value by about 10$\%$ (see text for explanation), the measured and computed anti-symmetric profiles are in the same orders of magnitude.  
 }
 \label{fig3}
 \end{figure}

We compute the Compton profile in the magnetic ground state of LiNiPO$_4$ along the three Cartesian directions. In agreement with the expectation from symmetry arguments, we find an anti-symmetric component in the Compton profile only along the $y$ direction, which is the direction of the toroidal moment $t_y$, in momentum space. Our calculated total Compton profile $J$ and the anti-symmetric part, $J^a$, are depicted in Fig. \ref{fig2}. Note that the anti-symmetric contribution has been multiplied by $10^3$. We can see from this figure that the anti-symmetric part of the profile is approximately four orders of magnitude smaller than the corresponding symmetric part of the profile. Although small, this signal should be at the limit of detectability, motivating the explicit measurements on LiNiPO$_4$, which we present in Section~\ref{exp}. 

Finally for this section, we analyze both parts of the profile to verify the zero-sum rules
\begin{eqnarray} \label{sumrule}
\int_{-\infty}^{\infty} p_y J^{s,a} (p_y) dp_y = 0 \quad . 
\end{eqnarray}
  This is a trivial condition for the symmetric part $J^s$ and is always satisfied because the integrand in Eq. \ref{sumrule} is by defnition always an odd function in $p_y$, resulting in a zero value upon integration. In contrast, for the anti-symmetric part, $J^a$, the zero-sum rule imposes strict conditions on the positive-$p_y$ and negative-$p_y$ halves of the profile separately \cite{Collins2016}, with
\begin{eqnarray}
\int_{-\infty}^{0} p_y J^{a} (p_y) dp_y = \int_{0}^{\infty} p_y J^{a} (p_y) dp_y = 0 \quad.
\end{eqnarray}
We find that our computed anti-symmetric Compton profile satisfies this condition to threee decimal places.

\subsection{Experimental results and comparison with theory} \label{exp}

In Fig. \ref{fig3} we show our measured total Compton profile (blue solid line with blue circles as data points) and the anti-symmetric signal (red circles with error bars, multiplied by $10^3$). 
The total profile is normalized so that the integral of the profile gives the total number of electrons in the unit cell of LiNiPO$_4$ with four formula units. The anti-symmetric profile is derived from the difference in Compton
profiles measured in the presence of magnetic fields with opposite directions (see Eq. \ref{Jas}). 
For comparison, the calculated profiles of Fig.~\ref{fig2}, after convolution with a Gaussian of full width at half maximum equal to our experimental resolution of 0.44 a.u., are also shown in Fig. \ref{fig3}. 
The blue dashed line and red solid line depict the calculated convoluted total and anti-symmetric profiles respectively.
As seen from Fig. \ref{fig3}, the peak in the calculated total profile is about 10$\%$ higher than the measured value. We also notice the skewness of the measured peak in the total Compton profile. We attribute both the skewness and the difference in measured and computed total profiles to multiple scattering effects, which are not considered here.
Multiple scattering, however, is not expected to have a significant affect on the anti-symmetric part of the profile.
Interestingly, we see that the data points corresponding to the measured anti-symmetric signal are consistent with our calculated anti-symmetric Compton profile, both being around four orders of magnitude smaller than the total profile. Frustratingly, however, both are at the level of the experimental noise (indicated by red error bars in Fig. \ref{fig3}), and we are not able to conclusively infer the existence of a non-zero antisymmetric  contribution. Our analysis suggests that the experiment came very close to measuring a signal of the calculated magnitude and that a modest increase in experimental signal to noise ratio would likely have yielded a positive result.

\section{Summary}  \label{sec3}

To summarize, we have studied the Compton profile in LiNiPO$_4$, the magnetic ground state of which allows for a net  toroidal moment $t_y$. Our density functional calculations show the existence of an anti-symmetric component in the Compton profile of LiNiPO$_4$ along the same $y$ direction in the momentum space as the toroidal moment, implicating anti-symmetric Compton scattering as a possible signature of a time-odd, parity-odd ME toroidal moment $t_y$ in the material. The calculated magnitude of the computed anti-symmetric component is small, however, with magnitude  $\sim 10^{-4}$ times the calculated total Compton profile. 

Our Compton scattering measurements on LiNiPO$_4$ also find a weak difference signal, consistent with the computed order of magnitude. Unfortunately, however, the weak signal is of the same order of magnitude as the statistical error, preventing us from conclusively determining an anti-symmetric profile. 

Our finding that the predicted and measured antisymmetric Compton response of LiNiPO$_4$ is exactly at the noise level motivates further research in two directions: First, materials with a larger antisymmetric Compton profile should be identified, so that existing experiments will provide an unambiguous signal. 
Second, since the experimental signal to noise ratio was limited by the detector count-rate capability, solid angle, incident beam flux and monochromator bandwidth, optimization of these factors, while practically challenging, could improve the signal to noise by an order of magnitude, allowing antisymmetric Compton profiles to be determined with ease.

\section*{Acknowledgements}
N. A. S. and S. B. were supported by the European Research Council (ERC) under the European Union’s Horizon 2020 research and innovation programme grant agreement No 810451 and by the ETH Zurich. Calculations were performed on the ETH Z\"urich Euler cluster. The synchrotron radiation experiments were performed at SPring-8 BL08W with the approval of the Japan Synchrotron Radiation Research Institute (JASRI) (Proposal No. 2018A1091).


 \bibliographystyle{apsrev4-1}
\bibliography{LNPO}

\begin{thebibliography}{23}%
\makeatletter
\providecommand \@ifxundefined [1]{%
 \@ifx{#1\undefined}
}%
\providecommand \@ifnum [1]{%
 \ifnum #1\expandafter \@firstoftwo
 \else \expandafter \@secondoftwo
 \fi
}%
\providecommand \@ifx [1]{%
 \ifx #1\expandafter \@firstoftwo
 \else \expandafter \@secondoftwo
 \fi
}%
\providecommand \natexlab [1]{#1}%
\providecommand \enquote  [1]{``#1''}%
\providecommand \bibnamefont  [1]{#1}%
\providecommand \bibfnamefont [1]{#1}%
\providecommand \citenamefont [1]{#1}%
\providecommand \href@noop [0]{\@secondoftwo}%
\providecommand \href [0]{\begingroup \@sanitize@url \@href}%
\providecommand \@href[1]{\@@startlink{#1}\@@href}%
\providecommand \@@href[1]{\endgroup#1\@@endlink}%
\providecommand \@sanitize@url [0]{\catcode `\\12\catcode `\$12\catcode
  `\&12\catcode `\#12\catcode `\^12\catcode `\_12\catcode `\%12\relax}%
\providecommand \@@startlink[1]{}%
\providecommand \@@endlink[0]{}%
\providecommand \url  [0]{\begingroup\@sanitize@url \@url }%
\providecommand \@url [1]{\endgroup\@href {#1}{\urlprefix }}%
\providecommand \urlprefix  [0]{URL }%
\providecommand \Eprint [0]{\href }%
\providecommand \doibase [0]{http://dx.doi.org/}%
\providecommand \selectlanguage [0]{\@gobble}%
\providecommand \bibinfo  [0]{\@secondoftwo}%
\providecommand \bibfield  [0]{\@secondoftwo}%
\providecommand \translation [1]{[#1]}%
\providecommand \BibitemOpen [0]{}%
\providecommand \bibitemStop [0]{}%
\providecommand \bibitemNoStop [0]{.\EOS\space}%
\providecommand \EOS [0]{\spacefactor3000\relax}%
\providecommand \BibitemShut  [1]{\csname bibitem#1\endcsname}%
\let\auto@bib@innerbib\@empty
\bibitem [{\citenamefont {Spaldin}\ \emph {et~al.}(2008)\citenamefont
  {Spaldin}, \citenamefont {Fiebig},\ and\ \citenamefont
  {Mostovoy}}]{Spaldin2008}%
  \BibitemOpen
  \bibfield  {author} {\bibinfo {author} {\bibfnamefont {N.~A.}\ \bibnamefont
  {Spaldin}}, \bibinfo {author} {\bibfnamefont {M.}~\bibnamefont {Fiebig}}, \
  and\ \bibinfo {author} {\bibfnamefont {M.}~\bibnamefont {Mostovoy}},\ }\href
  {\doibase 10.1088/0953-8984/20/43/434203} {\bibfield  {journal} {\bibinfo
  {journal} {Journal of Physics: Condensed Matter}\ }\textbf {\bibinfo {volume}
  {20}},\ \bibinfo {pages} {434203} (\bibinfo {year} {2008})}\BibitemShut
  {NoStop}%
\bibitem [{\citenamefont {Spaldin}\ \emph {et~al.}(2013)\citenamefont
  {Spaldin}, \citenamefont {Fechner}, \citenamefont {Bousquet}, \citenamefont
  {Balatsky},\ and\ \citenamefont {Nordstr\"om}}]{Spaldin2013}%
  \BibitemOpen
  \bibfield  {author} {\bibinfo {author} {\bibfnamefont {N.~A.}\ \bibnamefont
  {Spaldin}}, \bibinfo {author} {\bibfnamefont {M.}~\bibnamefont {Fechner}},
  \bibinfo {author} {\bibfnamefont {E.}~\bibnamefont {Bousquet}}, \bibinfo
  {author} {\bibfnamefont {A.}~\bibnamefont {Balatsky}}, \ and\ \bibinfo
  {author} {\bibfnamefont {L.}~\bibnamefont {Nordstr\"om}},\ }\href {\doibase
  10.1103/PhysRevB.88.094429} {\bibfield  {journal} {\bibinfo  {journal} {Phys.
  Rev. B}\ }\textbf {\bibinfo {volume} {88}},\ \bibinfo {pages} {094429}
  (\bibinfo {year} {2013})}\BibitemShut {NoStop}%
\bibitem [{\citenamefont {Spaldin}(2021)}]{Spaldin_2021}%
  \BibitemOpen
  \bibfield  {author} {\bibinfo {author} {\bibfnamefont {N.~A.}\ \bibnamefont
  {Spaldin}},\ }\href {\doibase 10.31857/S0044451021040027} {\bibfield
  {journal} {\bibinfo  {journal} {J. Exp. Theor. Phys}\ }\textbf {\bibinfo
  {volume} {159}},\ \bibinfo {pages} {594} (\bibinfo {year}
  {2021})}\BibitemShut {NoStop}%
\bibitem [{\citenamefont {Schmid}(2003)}]{Schmid}%
  \BibitemOpen
  \bibfield  {author} {\bibinfo {author} {\bibfnamefont {H.}~\bibnamefont
  {Schmid}},\ }\href@noop {} {\emph {\bibinfo {title} {Introduction to Complex
  Mediums for Optics and Electromagnetics}}}\ (\bibinfo  {publisher} {edited by
  W. S. Weiglhoger and A. Lakhtakia (SPIE Press, Bellingham, WA), pp.
  167–195},\ \bibinfo {year} {2003})\BibitemShut {NoStop}%
\bibitem [{\citenamefont {Van~Aken}\ \emph {et~al.}(2007)\citenamefont
  {Van~Aken}, \citenamefont {Rivera}, \citenamefont {Schmid},\ and\
  \citenamefont {Fiebig}}]{Aken}%
  \BibitemOpen
  \bibfield  {author} {\bibinfo {author} {\bibfnamefont {B.~B.}\ \bibnamefont
  {Van~Aken}}, \bibinfo {author} {\bibfnamefont {J.-P.}\ \bibnamefont
  {Rivera}}, \bibinfo {author} {\bibfnamefont {H.}~\bibnamefont {Schmid}}, \
  and\ \bibinfo {author} {\bibfnamefont {M.}~\bibnamefont {Fiebig}},\ }\href
  {\doibase 10.1038/nature06139} {\bibfield  {journal} {\bibinfo  {journal}
  {Nature}\ }\textbf {\bibinfo {volume} {449}},\ \bibinfo {pages} {702}
  (\bibinfo {year} {2007})}\BibitemShut {NoStop}%
\bibitem [{\citenamefont {Arima}\ \emph {et~al.}(2005)\citenamefont {Arima},
  \citenamefont {Jung}, \citenamefont {Matsubara}, \citenamefont {Kubota},
  \citenamefont {He}, \citenamefont {Kaneko},\ and\ \citenamefont
  {Tokura}}]{Arima2005}%
  \BibitemOpen
  \bibfield  {author} {\bibinfo {author} {\bibfnamefont {T.-h.}\ \bibnamefont
  {Arima}}, \bibinfo {author} {\bibfnamefont {J.-H.}\ \bibnamefont {Jung}},
  \bibinfo {author} {\bibfnamefont {M.}~\bibnamefont {Matsubara}}, \bibinfo
  {author} {\bibfnamefont {M.}~\bibnamefont {Kubota}}, \bibinfo {author}
  {\bibfnamefont {J.-P.}\ \bibnamefont {He}}, \bibinfo {author} {\bibfnamefont
  {Y.}~\bibnamefont {Kaneko}}, \ and\ \bibinfo {author} {\bibfnamefont
  {Y.}~\bibnamefont {Tokura}},\ }\href {\doibase 10.1143/JPSJ.74.1419}
  {\bibfield  {journal} {\bibinfo  {journal} {Journal of the Physical Society
  of Japan}\ }\textbf {\bibinfo {volume} {74}},\ \bibinfo {pages} {1419}
  (\bibinfo {year} {2005})},\ \Eprint
  {http://arxiv.org/abs/https://doi.org/10.1143/JPSJ.74.1419}
  {https://doi.org/10.1143/JPSJ.74.1419} \BibitemShut {NoStop}%
\bibitem [{\citenamefont {Staub}\ \emph {et~al.}(2009)\citenamefont {Staub},
  \citenamefont {Bodenthin}, \citenamefont {Piamonteze}, \citenamefont
  {Garc\'{\i}a-Fern\'andez}, \citenamefont {Scagnoli}, \citenamefont
  {Garganourakis}, \citenamefont {Koohpayeh}, \citenamefont {Fort},\ and\
  \citenamefont {Lovesey}}]{Staub2009}%
  \BibitemOpen
  \bibfield  {author} {\bibinfo {author} {\bibfnamefont {U.}~\bibnamefont
  {Staub}}, \bibinfo {author} {\bibfnamefont {Y.}~\bibnamefont {Bodenthin}},
  \bibinfo {author} {\bibfnamefont {C.}~\bibnamefont {Piamonteze}}, \bibinfo
  {author} {\bibfnamefont {M.}~\bibnamefont {Garc\'{\i}a-Fern\'andez}},
  \bibinfo {author} {\bibfnamefont {V.}~\bibnamefont {Scagnoli}}, \bibinfo
  {author} {\bibfnamefont {M.}~\bibnamefont {Garganourakis}}, \bibinfo {author}
  {\bibfnamefont {S.}~\bibnamefont {Koohpayeh}}, \bibinfo {author}
  {\bibfnamefont {D.}~\bibnamefont {Fort}}, \ and\ \bibinfo {author}
  {\bibfnamefont {S.~W.}\ \bibnamefont {Lovesey}},\ }\href {\doibase
  10.1103/PhysRevB.80.140410} {\bibfield  {journal} {\bibinfo  {journal} {Phys.
  Rev. B}\ }\textbf {\bibinfo {volume} {80}},\ \bibinfo {pages} {140410}
  (\bibinfo {year} {2009})}\BibitemShut {NoStop}%
\bibitem [{\citenamefont {Lovesey}(2014)}]{Lovesey}%
  \BibitemOpen
  \bibfield  {author} {\bibinfo {author} {\bibfnamefont {S.~W.}\ \bibnamefont
  {Lovesey}},\ }\href {\doibase 10.1088/0953-8984/26/35/356001} {\bibfield
  {journal} {\bibinfo  {journal} {Journal of Physics: Condensed Matter}\
  }\textbf {\bibinfo {volume} {26}},\ \bibinfo {pages} {356001} (\bibinfo
  {year} {2014})}\BibitemShut {NoStop}%
\bibitem [{\citenamefont {Staub}\ \emph {et~al.}(2010)\citenamefont {Staub},
  \citenamefont {Bodenthin}, \citenamefont {Piamonteze}, \citenamefont
  {Collins}, \citenamefont {Koohpayeh}, \citenamefont {Fort},\ and\
  \citenamefont {Lovesey}}]{Staub2010}%
  \BibitemOpen
  \bibfield  {author} {\bibinfo {author} {\bibfnamefont {U.}~\bibnamefont
  {Staub}}, \bibinfo {author} {\bibfnamefont {Y.}~\bibnamefont {Bodenthin}},
  \bibinfo {author} {\bibfnamefont {C.}~\bibnamefont {Piamonteze}}, \bibinfo
  {author} {\bibfnamefont {S.~P.}\ \bibnamefont {Collins}}, \bibinfo {author}
  {\bibfnamefont {S.}~\bibnamefont {Koohpayeh}}, \bibinfo {author}
  {\bibfnamefont {D.}~\bibnamefont {Fort}}, \ and\ \bibinfo {author}
  {\bibfnamefont {S.~W.}\ \bibnamefont {Lovesey}},\ }\href {\doibase
  10.1103/PhysRevB.82.104411} {\bibfield  {journal} {\bibinfo  {journal} {Phys.
  Rev. B}\ }\textbf {\bibinfo {volume} {82}},\ \bibinfo {pages} {104411}
  (\bibinfo {year} {2010})}\BibitemShut {NoStop}%
\bibitem [{\citenamefont {Scagnoli}\ \emph {et~al.}(2011)\citenamefont
  {Scagnoli}, \citenamefont {Staub}, \citenamefont {Bodenthin}, \citenamefont
  {de~Souza}, \citenamefont {Garc{\'\i}a-Fern{\'a}ndez}, \citenamefont
  {Garganourakis}, \citenamefont {Boothroyd}, \citenamefont {Prabhakaran},\
  and\ \citenamefont {Lovesey}}]{Staub}%
  \BibitemOpen
  \bibfield  {author} {\bibinfo {author} {\bibfnamefont {V.}~\bibnamefont
  {Scagnoli}}, \bibinfo {author} {\bibfnamefont {U.}~\bibnamefont {Staub}},
  \bibinfo {author} {\bibfnamefont {Y.}~\bibnamefont {Bodenthin}}, \bibinfo
  {author} {\bibfnamefont {R.~A.}\ \bibnamefont {de~Souza}}, \bibinfo {author}
  {\bibfnamefont {M.}~\bibnamefont {Garc{\'\i}a-Fern{\'a}ndez}}, \bibinfo
  {author} {\bibfnamefont {M.}~\bibnamefont {Garganourakis}}, \bibinfo {author}
  {\bibfnamefont {A.~T.}\ \bibnamefont {Boothroyd}}, \bibinfo {author}
  {\bibfnamefont {D.}~\bibnamefont {Prabhakaran}}, \ and\ \bibinfo {author}
  {\bibfnamefont {S.~W.}\ \bibnamefont {Lovesey}},\ }\href {\doibase
  10.1126/science.1201061} {\bibfield  {journal} {\bibinfo  {journal}
  {Science}\ }\textbf {\bibinfo {volume} {332}},\ \bibinfo {pages} {696}
  (\bibinfo {year} {2011})}\BibitemShut {NoStop}%
\bibitem [{\citenamefont {Kubota}\ \emph {et~al.}(2004)\citenamefont {Kubota},
  \citenamefont {Arima}, \citenamefont {Kaneko}, \citenamefont {He},
  \citenamefont {Yu},\ and\ \citenamefont {Tokura}}]{Kubota}%
  \BibitemOpen
  \bibfield  {author} {\bibinfo {author} {\bibfnamefont {M.}~\bibnamefont
  {Kubota}}, \bibinfo {author} {\bibfnamefont {T.}~\bibnamefont {Arima}},
  \bibinfo {author} {\bibfnamefont {Y.}~\bibnamefont {Kaneko}}, \bibinfo
  {author} {\bibfnamefont {J.~P.}\ \bibnamefont {He}}, \bibinfo {author}
  {\bibfnamefont {X.~Z.}\ \bibnamefont {Yu}}, \ and\ \bibinfo {author}
  {\bibfnamefont {Y.}~\bibnamefont {Tokura}},\ }\href {\doibase
  10.1103/PhysRevLett.92.137401} {\bibfield  {journal} {\bibinfo  {journal}
  {Phys. Rev. Lett.}\ }\textbf {\bibinfo {volume} {92}},\ \bibinfo {pages}
  {137401} (\bibinfo {year} {2004})}\BibitemShut {NoStop}%
\bibitem [{\citenamefont {Sessoli}\ \emph {et~al.}(2015)\citenamefont
  {Sessoli}, \citenamefont {Boulon}, \citenamefont {Caneschi}, \citenamefont
  {Mannini}, \citenamefont {Poggini}, \citenamefont {Wilhelm},\ and\
  \citenamefont {Rogalev}}]{Sessoli}%
  \BibitemOpen
  \bibfield  {author} {\bibinfo {author} {\bibfnamefont {R.}~\bibnamefont
  {Sessoli}}, \bibinfo {author} {\bibfnamefont {M.-E.}\ \bibnamefont {Boulon}},
  \bibinfo {author} {\bibfnamefont {A.}~\bibnamefont {Caneschi}}, \bibinfo
  {author} {\bibfnamefont {M.}~\bibnamefont {Mannini}}, \bibinfo {author}
  {\bibfnamefont {L.}~\bibnamefont {Poggini}}, \bibinfo {author} {\bibfnamefont
  {F.}~\bibnamefont {Wilhelm}}, \ and\ \bibinfo {author} {\bibfnamefont
  {A.}~\bibnamefont {Rogalev}},\ }\href {\doibase 10.1038/nphys3152} {\bibfield
   {journal} {\bibinfo  {journal} {Nature physics}\ }\textbf {\bibinfo {volume}
  {11}},\ \bibinfo {pages} {69} (\bibinfo {year} {2015})}\BibitemShut {NoStop}%
\bibitem [{\citenamefont {Jung}\ \emph {et~al.}(2004)\citenamefont {Jung},
  \citenamefont {Matsubara}, \citenamefont {Arima}, \citenamefont {He},
  \citenamefont {Kaneko},\ and\ \citenamefont {Tokura}}]{Jung2004}%
  \BibitemOpen
  \bibfield  {author} {\bibinfo {author} {\bibfnamefont {J.~H.}\ \bibnamefont
  {Jung}}, \bibinfo {author} {\bibfnamefont {M.}~\bibnamefont {Matsubara}},
  \bibinfo {author} {\bibfnamefont {T.}~\bibnamefont {Arima}}, \bibinfo
  {author} {\bibfnamefont {J.~P.}\ \bibnamefont {He}}, \bibinfo {author}
  {\bibfnamefont {Y.}~\bibnamefont {Kaneko}}, \ and\ \bibinfo {author}
  {\bibfnamefont {Y.}~\bibnamefont {Tokura}},\ }\href {\doibase
  10.1103/PhysRevLett.93.037403} {\bibfield  {journal} {\bibinfo  {journal}
  {Phys. Rev. Lett.}\ }\textbf {\bibinfo {volume} {93}},\ \bibinfo {pages}
  {037403} (\bibinfo {year} {2004})}\BibitemShut {NoStop}%
\bibitem [{\citenamefont {Ederer}\ and\ \citenamefont
  {Spaldin}(2007)}]{Claude2007}%
  \BibitemOpen
  \bibfield  {author} {\bibinfo {author} {\bibfnamefont {C.}~\bibnamefont
  {Ederer}}\ and\ \bibinfo {author} {\bibfnamefont {N.~A.}\ \bibnamefont
  {Spaldin}},\ }\href {\doibase 10.1103/PhysRevB.76.214404} {\bibfield
  {journal} {\bibinfo  {journal} {Phys. Rev. B}\ }\textbf {\bibinfo {volume}
  {76}},\ \bibinfo {pages} {214404} (\bibinfo {year} {2007})}\BibitemShut
  {NoStop}%
\bibitem [{\citenamefont {Collins}\ \emph {et~al.}(2016)\citenamefont
  {Collins}, \citenamefont {Laundy}, \citenamefont {Connolley}, \citenamefont
  {van~der Laan}, \citenamefont {Fabrizi}, \citenamefont {Janssen},
  \citenamefont {Cooper}, \citenamefont {Ebert},\ and\ \citenamefont
  {Mankovsky}}]{Collins2016}%
  \BibitemOpen
  \bibfield  {author} {\bibinfo {author} {\bibfnamefont {S.~P.}\ \bibnamefont
  {Collins}}, \bibinfo {author} {\bibfnamefont {D.}~\bibnamefont {Laundy}},
  \bibinfo {author} {\bibfnamefont {T.}~\bibnamefont {Connolley}}, \bibinfo
  {author} {\bibfnamefont {G.}~\bibnamefont {van~der Laan}}, \bibinfo {author}
  {\bibfnamefont {F.}~\bibnamefont {Fabrizi}}, \bibinfo {author} {\bibfnamefont
  {O.}~\bibnamefont {Janssen}}, \bibinfo {author} {\bibfnamefont {M.~J.}\
  \bibnamefont {Cooper}}, \bibinfo {author} {\bibfnamefont {H.}~\bibnamefont
  {Ebert}}, \ and\ \bibinfo {author} {\bibfnamefont {S.}~\bibnamefont
  {Mankovsky}},\ }\href {\doibase 10.1107/S2053273316000863} {\bibfield
  {journal} {\bibinfo  {journal} {Acta Crystallographica Section A}\ }\textbf
  {\bibinfo {volume} {72}},\ \bibinfo {pages} {197} (\bibinfo {year}
  {2016})}\BibitemShut {NoStop}%
\bibitem [{\citenamefont {Mercier}\ and\ \citenamefont
  {Bauer}(1968)}]{Mercier}%
  \BibitemOpen
  \bibfield  {author} {\bibinfo {author} {\bibfnamefont {M.}~\bibnamefont
  {Mercier}}\ and\ \bibinfo {author} {\bibfnamefont {P.}~\bibnamefont
  {Bauer}},\ }\href@noop {} {\bibfield  {journal} {\bibinfo  {journal} {C. R.
  Acad. Sci. Paris}\ }\textbf {\bibinfo {volume} {267}},\ \bibinfo {pages}
  {465} (\bibinfo {year} {1968})}\BibitemShut {NoStop}%
\bibitem [{\citenamefont {Jensen}\ \emph {et~al.}(2009)\citenamefont {Jensen},
  \citenamefont {Christensen}, \citenamefont {Kenzelmann}, \citenamefont
  {R\o{}nnow}, \citenamefont {Niedermayer}, \citenamefont {Andersen},
  \citenamefont {Lefmann}, \citenamefont {Schefer}, \citenamefont
  {v.~Zimmermann}, \citenamefont {Li}, \citenamefont {Zarestky},\ and\
  \citenamefont {Vaknin}}]{Jensen}%
  \BibitemOpen
  \bibfield  {author} {\bibinfo {author} {\bibfnamefont {T.~B.~S.}\
  \bibnamefont {Jensen}}, \bibinfo {author} {\bibfnamefont {N.~B.}\
  \bibnamefont {Christensen}}, \bibinfo {author} {\bibfnamefont
  {M.}~\bibnamefont {Kenzelmann}}, \bibinfo {author} {\bibfnamefont {H.~M.}\
  \bibnamefont {R\o{}nnow}}, \bibinfo {author} {\bibfnamefont {C.}~\bibnamefont
  {Niedermayer}}, \bibinfo {author} {\bibfnamefont {N.~H.}\ \bibnamefont
  {Andersen}}, \bibinfo {author} {\bibfnamefont {K.}~\bibnamefont {Lefmann}},
  \bibinfo {author} {\bibfnamefont {J.}~\bibnamefont {Schefer}}, \bibinfo
  {author} {\bibfnamefont {M.}~\bibnamefont {v.~Zimmermann}}, \bibinfo {author}
  {\bibfnamefont {J.}~\bibnamefont {Li}}, \bibinfo {author} {\bibfnamefont
  {J.~L.}\ \bibnamefont {Zarestky}}, \ and\ \bibinfo {author} {\bibfnamefont
  {D.}~\bibnamefont {Vaknin}},\ }\href {\doibase 10.1103/PhysRevB.79.092412}
  {\bibfield  {journal} {\bibinfo  {journal} {Phys. Rev. B}\ }\textbf {\bibinfo
  {volume} {79}},\ \bibinfo {pages} {092412} (\bibinfo {year}
  {2009})}\BibitemShut {NoStop}%
\bibitem [{\citenamefont {Fogh}\ \emph {et~al.}(2020)\citenamefont {Fogh},
  \citenamefont {Kihara}, \citenamefont {Toft-Petersen}, \citenamefont
  {Bartkowiak}, \citenamefont {Narumi}, \citenamefont {Prokhnenko},
  \citenamefont {Miyake}, \citenamefont {Tokunaga}, \citenamefont {Oikawa},
  \citenamefont {S\o{}rensen}, \citenamefont {Dyrnum}, \citenamefont {Grimmer},
  \citenamefont {Nojiri},\ and\ \citenamefont {Christensen}}]{Exp2020}%
  \BibitemOpen
  \bibfield  {author} {\bibinfo {author} {\bibfnamefont {E.}~\bibnamefont
  {Fogh}}, \bibinfo {author} {\bibfnamefont {T.}~\bibnamefont {Kihara}},
  \bibinfo {author} {\bibfnamefont {R.}~\bibnamefont {Toft-Petersen}}, \bibinfo
  {author} {\bibfnamefont {M.}~\bibnamefont {Bartkowiak}}, \bibinfo {author}
  {\bibfnamefont {Y.}~\bibnamefont {Narumi}}, \bibinfo {author} {\bibfnamefont
  {O.}~\bibnamefont {Prokhnenko}}, \bibinfo {author} {\bibfnamefont
  {A.}~\bibnamefont {Miyake}}, \bibinfo {author} {\bibfnamefont
  {M.}~\bibnamefont {Tokunaga}}, \bibinfo {author} {\bibfnamefont
  {K.}~\bibnamefont {Oikawa}}, \bibinfo {author} {\bibfnamefont {M.~K.}\
  \bibnamefont {S\o{}rensen}}, \bibinfo {author} {\bibfnamefont {J.~C.}\
  \bibnamefont {Dyrnum}}, \bibinfo {author} {\bibfnamefont {H.}~\bibnamefont
  {Grimmer}}, \bibinfo {author} {\bibfnamefont {H.}~\bibnamefont {Nojiri}}, \
  and\ \bibinfo {author} {\bibfnamefont {N.~B.}\ \bibnamefont {Christensen}},\
  }\href {\doibase 10.1103/PhysRevB.101.024403} {\bibfield  {journal} {\bibinfo
   {journal} {Phys. Rev. B}\ }\textbf {\bibinfo {volume} {101}},\ \bibinfo
  {pages} {024403} (\bibinfo {year} {2020})}\BibitemShut {NoStop}%
\bibitem [{\citenamefont {Abrahams}\ and\ \citenamefont
  {Easson}(1993)}]{Abrahams}%
  \BibitemOpen
  \bibfield  {author} {\bibinfo {author} {\bibfnamefont {I.}~\bibnamefont
  {Abrahams}}\ and\ \bibinfo {author} {\bibfnamefont {K.~S.}\ \bibnamefont
  {Easson}},\ }\href {\doibase 10.1107/S0108270192013064} {\bibfield  {journal}
  {\bibinfo  {journal} {Acta Crystallographica Section C}\ }\textbf {\bibinfo
  {volume} {49}},\ \bibinfo {pages} {925} (\bibinfo {year} {1993})}\BibitemShut
  {NoStop}%
\bibitem [{cod()}]{code}%
  \BibitemOpen
  \href@noop {} {\enquote {\bibinfo {title} {{The Elk Code}},}\ }\bibinfo
  {howpublished} {\url{http://elk.sourceforge.net/}}\BibitemShut {NoStop}%
\bibitem [{\citenamefont {Ernsting}\ \emph {et~al.}(2014)\citenamefont
  {Ernsting}, \citenamefont {Billington}, \citenamefont {Haynes}, \citenamefont
  {Millichamp}, \citenamefont {Taylor}, \citenamefont {Duffy}, \citenamefont
  {Giblin}, \citenamefont {Dewhurst},\ and\ \citenamefont {Dugdale}}]{elk}%
  \BibitemOpen
  \bibfield  {author} {\bibinfo {author} {\bibfnamefont {D.}~\bibnamefont
  {Ernsting}}, \bibinfo {author} {\bibfnamefont {D.}~\bibnamefont
  {Billington}}, \bibinfo {author} {\bibfnamefont {T.~D.}\ \bibnamefont
  {Haynes}}, \bibinfo {author} {\bibfnamefont {T.~E.}\ \bibnamefont
  {Millichamp}}, \bibinfo {author} {\bibfnamefont {J.~W.}\ \bibnamefont
  {Taylor}}, \bibinfo {author} {\bibfnamefont {J.~A.}\ \bibnamefont {Duffy}},
  \bibinfo {author} {\bibfnamefont {S.~R.}\ \bibnamefont {Giblin}}, \bibinfo
  {author} {\bibfnamefont {J.~K.}\ \bibnamefont {Dewhurst}}, \ and\ \bibinfo
  {author} {\bibfnamefont {S.~B.}\ \bibnamefont {Dugdale}},\ }\href {\doibase
  10.1088/0953-8984/26/49/495501} {\bibfield  {journal} {\bibinfo  {journal}
  {Journal of Physics: Condensed Matter}\ }\textbf {\bibinfo {volume} {26}},\
  \bibinfo {pages} {495501} (\bibinfo {year} {2014})}\BibitemShut {NoStop}%
\bibitem [{\citenamefont {Biggs}\ \emph {et~al.}(1975)\citenamefont {Biggs},
  \citenamefont {Mendelsohn},\ and\ \citenamefont {Mann}}]{Biggs}%
  \BibitemOpen
  \bibfield  {author} {\bibinfo {author} {\bibfnamefont {F.}~\bibnamefont
  {Biggs}}, \bibinfo {author} {\bibfnamefont {L.}~\bibnamefont {Mendelsohn}}, \
  and\ \bibinfo {author} {\bibfnamefont {J.}~\bibnamefont {Mann}},\ }\href
  {\doibase https://doi.org/10.1016/0092-640X(75)90030-3} {\bibfield  {journal}
  {\bibinfo  {journal} {Atomic Data and Nuclear Data Tables}\ }\textbf
  {\bibinfo {volume} {16}},\ \bibinfo {pages} {201} (\bibinfo {year}
  {1975})}\BibitemShut {NoStop}%
\bibitem [{\citenamefont {Cooper}(1985)}]{Cooper1985}%
  \BibitemOpen
  \bibfield  {author} {\bibinfo {author} {\bibfnamefont {M.~J.}\ \bibnamefont
  {Cooper}},\ }\href {\doibase 10.1088/0034-4885/48/4/001} {\bibfield
  {journal} {\bibinfo  {journal} {Reports on Progress in Physics}\ }\textbf
  {\bibinfo {volume} {48}},\ \bibinfo {pages} {415} (\bibinfo {year}
  {1985})}\BibitemShut {NoStop}%
\end{thebibliography}%

\end{document}